# Impact of Non-Linear High-Power Amplifiers on Cooperative Relaying Systems

Elyes Balti, *Student Member, IEEE*, and Mohsen Guizani, *Fellow, IEEE*

*Abstract*—In this paper, we investigate the impact of the high-power amplifier non-linear distortion on multiple relay systems by introducing the soft envelope limiter, traveling wave tube amplifier, and solid-state power amplifier to the relays. The system employs amplify-and-forward either fixed or variable gain relaying and uses the opportunistic relay selection with outdated channel state information to select the best relay. The results show that the performance loss is small at low rates; however, it is significant for high rates. In particular, the outage probability and the bit error rate are saturated by an irreducible floor at high rates. The same analysis is pursued for the capacity and shows that it is saturated by a detrimental ceiling as the average signal-to-noise ratio becomes higher. This result contrasts the case of the ideal hardware where the capacity grows indefinitely. Moreover, the results show that the capacity ceiling is proportional to the impairment's parameter and for some special cases the impaired systems practically operate in acceptable conditions. Closed-forms and high SNR asymptotes of the outage probability, the bit error rate, and the capacity are derived. Finally, analytical expressions are validated by the Monte Carlo simulation.

*Index Terms*—Soft envelope limiter, traveling wave tube amplifier, solid state power amplifier, amplify-and-forward, imperfect CSI, opportunistic relay selection.

## I. INTRODUCTION

COOPERATIVE relaying-assisted communication is the corner-stone of the next generation wireless communication systems because of numerous advantages, such as coverage extension, reliability, uniform quality of service (QoS) [1], spatial diversity gain and hostpot throughput improvement [2]–[6]. Consequently, future mobile broadband networks such as 3GPP LTE-Advanced, IEEE 802.16m and IEEE 802.16j are expected to support communications based on relaying. Based on that, relaying networks have gained enormous attention over the recent decade both in industry and academia [7]–[10].

In most networking systems, the relaying technique is achieved in two steps. In the first step, a source ($S$) transmits the signal and all the relays are sensing. In the second time slot, the relays cooperatively transmit the information symbols to the destination ($D$). There are many relaying techniques, but the most commonly used are Amplify-and-Forward (AF) [11]–[14], Decode-and-Forward (DF) [15]–[17] and Quantize-and-Encode/Forward [18], [19]. The benefits of cooperative relaying come at the expense of the low signal coverage in farther areas. In fact, some cellular areas suffer from low coverage and power outage and it has been shown that an efficient way to increase the coverage reliability and the network scalability is to implement a set of relays along the path between the base station and the farthest areas. Furthermore, the inefficient utilization of the spectrum can be reduced by using relay selection protocols. These protocols state that a single relay is selected following some specific rules to forward the signal to $D$.

### A. Literature Review

In the literature, there are many relay selection protocols, but the most popular are partial relay selection (PRS) and opportunistic relay selection (ORS) [20]–[23]. For the PRS, the selection is achieved based on the channel state information (CSI) of either the first or the second hop. However, ORS requires the knowledge of the CSI of the overall channels. Further details about this protocol will be given later in the next section. Although the PRS has low complexity, short network delay and low power consumption, ORS is known to be more efficient specifically, the signal outage, error performance and system capacity are better [21], [22]. In addition, the feedback signals between $S$, relays and $D$ are slowly propagating, so it is important to take into account this delay and consider outdated CSI rather than perfect channel estimation during the relay selection. Moreover, the outdated CSI can be considered for the amplification gain at the relay as well. This point will be detailed further in the section of the system model.

The vast majority of previous work assumed relaying system with ideal transceivers [24]–[28]. However, in practice the transceivers are susceptible to many types of imperfections such as HPA non-linearities, In phase and Quadrature phase (I/Q) imbalance, phase noise, DC offset [29]–[32]. Schenk *et al.* [33] have considered I/Q imbalance and proved that this impairment attenuates the magnitude of the signal. Furthermore, Maletic *et al.* [34] characterized the effect of non-linear HPA and they demonstrated that the system performance such as the outage probability, BER and the ergodic capacity deteriorated compared to the linear HPA. As long as the impairment becomes more severe, an irreducible floor is created that it cannot be crossed by increasing the average transmitted power [35], [36].

Related work to cooperative relaying communication are prominent in the literature. In fact, Bjornson *et al.* [35] have considered a dual-hop system with single relay employing AF and DF relaying schemes wherein the source and the relay both suffer from aggregate hardware impairment. This work quantified the impacts of the impairments on the outage probability and the ergodic capacity. They proved that the capacity is finite and limited by a hardware ceiling and they also showed that DF is more resilient to the impairments than AF. The same research group also considered a two-way relaying under the presence of relay transceiver hardware impairments to prove that the outage probability and the symbol error probability are saturated by irreducible floors created by the hardware impairments. In the same context, Studer *et al.* [31] considered MIMO transmission with residual transmit-RF impairments wherein they proposed Tx-noise whitening technique to mitigate the performance loss. Moreover, Qi and Aïssa [37] provided a framework analysis of the compensation of the power amplifier non-linearity in MIMO transmit diversity systems wherein they derived the expressions of the total degradation, the symbol error rate and the system capacity. Advanced research attempts [38], [39] have considered mixed RF and FSO (Free-space optic) relaying system suffering from aggregate hardware impairments where the RF channels experience Rayleigh fading and the FSO channels are subject to Gamma-Gamma and Double Weibull fading, respectively. Furthermore, additional work have focused more on investigating the cooperative diversity of multiple relay systems but assuming ideal hardware. In fact, [21] and [22] have proposed dual-hop multiple relay systems with PRS and ORS protocols with outdated CSI wherein they derived the closed-forms of the outage probability, the BER and they also provided the diversity and the coding gains of the proposed systems. Although [21] and [22] came up with novel expressions, they neglected the impact of the impairments. Given that they assumed that such systems are promising for the advancement of future wireless communications since they are of high rate and hence, the assumption of neglecting the impairments cannot hold in this situation. To address this shortcoming, we keep the same configuration of the proposed system of [21] and [22] but we introduce hardware impairments in the relays. Such impairments' models presented by this work are detailed in the next subsection.

### B. HPA Overview

The origin of the HPA comes from the fact that the relaying amplification is not linear which creates a non-linear distortion that severely degrades the quality of the signal. In practice, there is a finite peak level for which any power amplifier can produce an output power without exceeding that power constraint. This peak constraint is primarily amplifier-dependent and varies within a given bounded range. If the amplifier is unable to provide the required power, a non-linear distortion over the peak is introduced and such phenomenon is called clipping (clipping factor) of the power amplifier.

The HPA can be classified into memoryless and with memory. In fact, the HPA is said to be memoryless or frequency-independent if its frequency response is constant over the operating frequencies range. In this case, the HPA is fully characterized by the famous characteristics AM/AM (amplitude to amplitude conversion) and AM/PM (amplitude to phase conversion). AM/AM and AM/PM will be given in more details in Section II-C. On the other side, if the frequency characteristics totally depend on either the frequency components or the thermal phenomena, the HPA is said to be with memory [40]. Such system can be characterized by realistic memory models, such as the Volterra, Wiener, Hammerstein, Wiener-Hammerstein and memory polynomial models [41].

In practice, there are various models of memoryless HPA but the most commonly known are Soft Envelope Limiter (SEL), Traveling Wave Tube Amplifier (TWTA) and Solid State Power Amplifier (SSPA) or also called the Rapp model [42]–[48]. The SEL is typically used to model a HPA with a perfect predistortion system while the TWTA has been basically employed to model the non-linearities impact in an OFDM system. In addition, SSPA is characterized by a smoothness factor to control the transition between the saturation and the linear ranges. This model eventually introduces a linear characteristic for low amplitudes of the input signal and then it is limited by an output saturation level. For larger values of the smoothness factor, SSPA practically converges to the SEL model.

### C. Contribution

In this paper, we introduce three models of HPA non-linearities at the relays, which are SEL, TWTA and SSPA [34], [49]. Then we will study the effect of the relay saturation on the outage probability, the average BER and the ergodic capacity under different relaying schemes. These relaying modes are fixed gain (FG), variable gain (VG) version I (VGI) and version II (VGII). Note that the first version of the variable gain scheme is based on calculating the amplification gain of the instantaneous CSI feedbacks between $S$, relays and $D$. The signal amplification will be based on this outdated CSI. For the second version of (VG), the relays are supposed to have an updated version of the CSI information to compute the amplification gain. To the best of our knowledge, this is the first work elaborating on a global framework analysis of multiple relays under the effect of various models of HPA non-linear distortion. We will show that both the outage and the error performances are saturated by inevitable floors while the system capacity is limited by a finite ceiling. For some special cases, we will show that the system can operate in acceptable conditions with the presence of the hardware impairments.

This work makes the following contributions:

- Present a detailed description of the system model and the relay selection protocol.
- Provide an analytical framework of the impairments and how to convert the non-linear distortion into a linear impact on the system using the Bussgang linearization theory.
- Present the statistics of the channels in terms of the high order moment, the probability density function (PDF) and the cumulative distribution function (CDF).

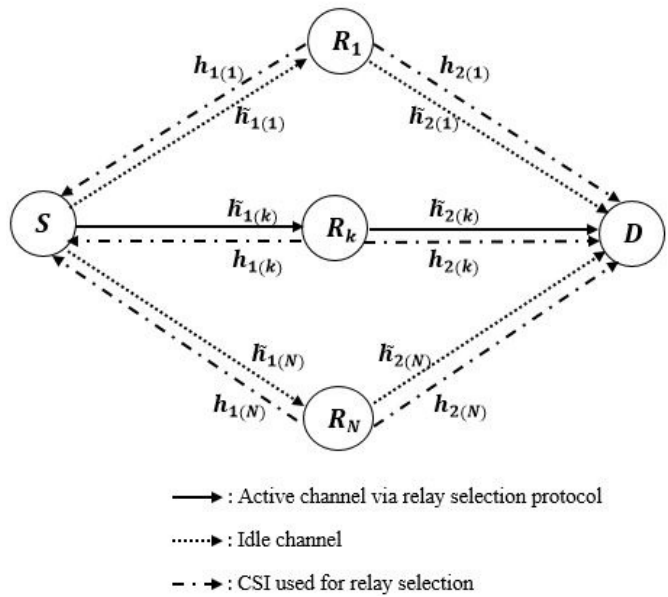

Fig. 1. Dual-hop cooperative relaying system.

- Once obtaining the signal-to-noise-plus-distortion-ratio (SNDR), which is a measure of the degradation of the signal by unwanted or extraneous signals including noise and distortion, we will derive the closed-forms of the outage probability, BER and the ergodic capacity for FG, VGI and VGII.
- Finally, to obtain further insights on the proposed system, we derive asymptotic expressions of the outage probability and BER at high signal-to-noise-ratio (SNR) regime. Capitalizing on these asymptotes, we derive the diversity gain of the proposed system.

### D. Structure

This paper is organized as follows: Section II describes the system model and the impairment types. The outage probability analysis is provided in Section III while the BER analysis is given in Section IV. Analytical and numerical results are detailed in Section V. Concluding remarks and future directions are presented in the final Section.

## II. System Model

The system is composed of a source $S$, destination $D$ and $N$ parallel relays $R_n, n = 1,.., N$ wirelessly connected to $S$ and $D$ as shown in Fig. 1. The channels of the first and the second hops are symmetric, independent and indentically distributed following the Rayleigh distribution.

### A. CSI Model

As we mentioned earlier, we assumed an outdated CSI instead of a perfect one. In this case, the relay selection protocol is achieved based on a delayed version of the CSI and not on the current one due to the feedback delay. In this way, the outdated and the current channels gains are denoted by $\widetilde{h}$ and $h$, respectively. Hence, the outdated CSI between $S$ - $k$th relay and $k$th relay - $D$ are, respectively, modeled as follows:

$$\widetilde{h}_{1(k)} = \sqrt{\rho_1} h_{1(k)} + \sqrt{1 - \rho_1} w_{1(k)} \qquad (1)$$

and

$$\widetilde{h}_{2(k)} = \sqrt{\rho_2} h_{2(k)} + \sqrt{1 - \rho_2} w_{2(k)} \qquad (2)$$

where $w_{1(k)}$ and $w_{2(k)}$ are two random variables that, respectively, follow the circularly symmetric complex Gaussian distribution with the same variances of the channels' gains $h_{1(k)}$ and $h_{2(k)}$. The time correlation coefficients $\rho_1$ and $\rho_2$ are between the channels $h_1 - \widetilde{h}_1$ and $h_2 - \widetilde{h}_2$, respectively. The coefficients $\rho_1$ and $\rho_2$ are given by the Jakes' autocorrelation model as follows [50]:

$$\rho_1 = J_0(2\pi f_{d,1} T_d) \qquad (3)$$

and

$$\rho_2 = J_0(2\pi f_{d,2} T_d) \qquad (4)$$

where $J_0(\cdot)$ is the zeroth order Bessel function of the first kind [51, eq. (8.411)], $T_d$ is the time delay between the current CSI and the delayed version and $f_d$ is the maximum Doppler frequency of the channels.

### B. Opportunistic Relay Selection

This protocol states that each relay should quantify its appropriateness as an active relay, using a function describing the link quality of the two hops. The first step is to select the minimum channel gains between two hops for each relay Eq. (5). Based on the first step, the relay of rank $k$ characterized by the strongest bottleneck is the one with the best overall path between $S$ and $D$ Eq. (6).

$$\gamma_i = \min(\gamma_{1(i)}, \gamma_{2(i)}) \qquad (5)$$

Then

$$k = \arg \max_i(\gamma_i) \qquad (6)$$

where $\gamma_{1(i)}, \gamma_{2(i)}$ are the instantaneous SNRs of the $i$th channel of the first and second hops, respectively.

Since the relays operate in a half-duplex mode, the best relay is not always available and so the control unit will select the next best available relay.

### C. HPA Non-Linearities Model

We assume that the relays are subject to HPA non-linearities. For a given transmission, the selected relay receives the signal $y_{1(k)}$ from $S$ and then amplifies it by the factor gain $G$. This amplification takes place in two time slots. In the first phase, the gain $G$ is applied to the received signal as follows:

$$\phi_k = G y_{1(k)} \qquad (7)$$

In the second phase, the output signal $\phi_k$ passes through a non-linear circuit as follows:

$$\psi_k = f(\phi_k) \qquad (8)$$

where $f(\cdot)$ is the function of amplitude and phase of the non-linear circuit. In addition, we assume that the relays power amplifiers are memoryless. A given memoryless power amplifier is characterized by both AM/AM and AM/PM. The signal at the output of the non-linear circuit is given by [37]:

$$\psi_k = F_a(\phi_k) \; \exp(j(\arg(\phi_k) + F_p(\phi_k)))) \quad (9)$$

where $\arg(\phi_k)$ is the phase of the complex signal $\phi_k$ and $F_a(\cdot)$, $F_p(\cdot)$ are the characteristic functions AM/AM, AM/PM, respectively.

*1) SEL:* This type of impairment is suitable to model a HPA with perfect predistortion system. The characteristic functions of SEL are expressed as follows [42]:

$$F_a(\phi_k) = \begin{cases} |\phi_k|, & |\phi_k| \le A_{sat} \\ A_{sat}, & \text{otherwise} \end{cases} ; \quad F_p(\phi_k) = 0 \quad (10)$$

where $A_{sat}$ is the HPA input saturation amplitude.

*2) SSPA:* This impairment model, also called the Rapp model, was detailed in [52] and presents only the amplitude characteristic AM/AM. The functions are given by:

$$F_a(\phi_k) = |\phi_k| \left[ 1 + \left( \frac{|\phi_k|}{A_{sat}} \right)^{2\nu} \right]^{-\frac{1}{2\nu}} ; \quad F_p(\phi_k) = 0 \quad (11)$$

where $\nu$ is the smoothness factor that controls the transition from linear to saturation domain. As $\nu$ converges to infinity, SSPA effectively converges to the SEL model.

*3) TWTA:* This impairment is used to model the impact of non-linearities in OFDM systems [53], [54]. The characteristic functions of this model are given by:

$$F_a(\phi_k) = A_{sat}^2 \frac{|\phi_k|}{|\phi_k|^2 + A_{sat}^2}; \quad F_p(\phi_k) = \Phi_0 \frac{|\phi_k|^2}{|\phi_k|^2 + A_{sat}^2} \quad (12)$$

where $\Phi_0$ controls the maximum phase distortion.

In practice, to mitigate the impacts of the non-linear distortion, the HPA operates at an input back-off (IBO) from a given saturation level. In the literature, there have been many definitions of the IBO, but in this work, we will adopt the following definition:

$$\text{IBO} = 10 \log_{10} \left( \frac{A_{sat}^2}{\sigma^2} \right) \quad (13)$$

where $\sigma^2$ is the mean power of the signal at the output of the gain block. Fig. 2 presents the variations of the AM/AM with respect to the normalized input modulus for SEL, SSPA and TWTA.

### D. Bussgang Linearization Theory

This theory states that the output of the non-linear power amplifier circuit can be expressed in terms of a linear scale parameter $\delta$ of the input signal and a non-linear distortion $\tau$ which is uncorrelated with the input signal and distributed following the complex circular Gaussian random variable $\tau \curvearrowright \mathcal{CN}(0, \; \sigma_\tau^2)$. In this case, the characteristic function of the amplitude is given by:

$$\psi_k = \delta \phi_k + \tau \quad (14)$$

We can derive the expressions of $\delta$ and $\sigma_\tau^2$ following the two corollaries.

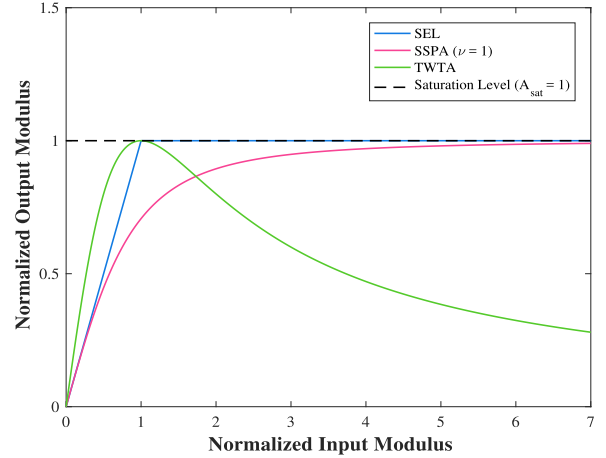

Fig. 2. AM/AM characteristics of SEL, SSPA and TWTA.

*Corollary 1:* The linear scale $\delta$ can be derived as follows:

$$\delta = \frac{\mathbb{E}\left[\phi_k^* \psi_k\right]}{\mathbb{E}\left[|\phi_k|^2\right]} \quad (15)$$

*Corollary 2:* The variance of the non-linear distortion is given by:

$$\sigma_\tau^2 = \mathbb{E}\left[|\psi_k|^2\right] - \delta \mathbb{E}\left[\phi_k \psi_k^*\right] \quad (16)$$

For the SEL model, $\delta$ and $\sigma_\tau^2$ can be expressed as follows [34, eq. (10)]:

$$\delta = 1 - \exp\left(-\frac{A_{sat}^2}{\sigma^2}\right) + \frac{\sqrt{\pi} A_{sat}}{2\sigma} \text{erfc}\left(-\frac{A_{sat}}{\sigma}\right)$$

$$\sigma_\tau^2 = \sigma^2 \left[ 1 - \exp\left(-\frac{A_{sat}^2}{\sigma^2}\right) - \delta^2 \right] \quad (17)$$

where $\text{erfc}(\cdot)$ is the complementary error function.

To simplify the calculation for the case of SSPA, we first assume that the smoothness factor ($\nu = 1$) and then we refer to [55] to derive the parameters as follows:

$$\delta = \frac{A_{sat}}{2\sigma} \left[ \frac{2A_{sat}}{\sigma} - \sqrt{\pi} \text{erfc}\left( \frac{A_{sat}}{\sigma} \right) \exp\left( \frac{A_{sat}^2}{\sigma^2} \right) \right.$$
$$\left. \times \left( \frac{2A_{sat}^2}{\sigma^2} - 1 \right) \right]$$

$$\sigma_\tau^2 = \sigma^2 \left[ \frac{A_{sat}^2}{\sigma^2} \left( 1 + \frac{A_{sat}^2}{\sigma^2} \exp\left( \frac{A_{sat}^2}{\sigma^2} \right) \text{Ei}\left( -\frac{A_{sat}^2}{\sigma^2} \right) \right) - \delta^2 \right] \quad (18)$$

where $\text{Ei}(\cdot)$ is the exponential integral function.

If the phase characteristic AM/PM is negligeable (i.e., $\Phi_0 \approx 0$), the impairment parameters $\delta$ and $\sigma^2$ for TWTA can be obtained by [34, eq. (11)]:

$$\delta = \frac{A_{sat}^2}{\sigma^2} \left[ 1 + \frac{A_{sat}^2}{\sigma^2} \exp\left( \frac{A_{sat}^2}{\sigma^2} \right) \text{Ei}\left( -\frac{A_{sat}^2}{\sigma^2} \right) \right]$$

$$\sigma_\tau^2 = -\frac{A_{sat}^4}{\sigma^2} \left[ \left( 1 + \frac{A_{sat}^2}{\sigma^2} \right) \exp\left( \frac{A_{sat}^2}{\sigma^2} \right) \text{Ei}\left( -\frac{A_{sat}^2}{\sigma^2} \right) + 1 \right]$$
$$- \sigma^2 \delta^2 \quad (19)$$

### E. Statistics of the Channels

Since the channels of the first hop experience Rayleigh fading with outdated CSI and the system employs the opportunistic relay selection protocol, the PDF of the SNR of the first hop of the $k$th channel is given by [22, eq. (21)]:

$$f_{\widetilde{\gamma}_{1(k)}}(x) = \frac{k}{\overline{\gamma}_1} \binom{N}{k} \sum_{n=0}^{k-1} \sum_{j=1}^{2} P_n Q_{n,j} \exp\left(-\frac{R_{n,j}x}{\overline{\gamma}_2}\right) \quad (20)$$

Due to the symmetry of the channels fading, the CDF of the second hop at the $k$th channel can be expressed as follows:

$$F_{\widetilde{\gamma}_{2(k)}}(x) = 1 - k\binom{N}{k} \sum_{m=0}^{k-1} \sum_{i=1}^{2} S_m T_{m,i} \exp\left(-\frac{U_{m,i}x}{\overline{\gamma}_1}\right) \quad (21)$$

where $P_n$, $S_m$, $Q_{n,j}$, $R_{n,j}$, $T_{m,i}$, $U_{m,i}$ and $\overline{\gamma}$ are defined by:

$$P_n = \frac{(-1)^n \binom{k-1}{n}}{1 + \frac{\overline{\gamma}_2}{\overline{\gamma}}(N-k+n)}; \quad S_m = \frac{(-1)^m \binom{k-1}{m}}{1 + \frac{\overline{\gamma}_1}{\overline{\gamma}}(N-k+m)}$$

$$Q_{n,1} = 1; \quad Q_{n,2} = \frac{(N-k+n)\overline{\gamma}_2}{\rho_1\overline{\gamma} + (1-\rho_1)(N-k+n+1)\overline{\gamma}_1}$$

$$R_{n,1} = 1; \quad R_{n,2} = \frac{(N-k+n+1)\overline{\gamma}_1}{\rho_1\overline{\gamma} + (1-\rho_1)(N-k+n+1)\overline{\gamma}_1}$$

$$T_{m,1} = 1; \quad T_{m,2} = \frac{(N-k+m)\overline{\gamma}_1}{(N-k+m+1)\overline{\gamma}_2}$$

$$U_{m,1} = \frac{\overline{\gamma}_1}{\overline{\gamma}_2}; \quad U_{m,2} = \frac{(N-k+m+1)\overline{\gamma}_2}{\rho_2\overline{\gamma} + (1-\rho_2)(N-k+m+1)\overline{\gamma}_2}$$

$$\overline{\gamma} = \frac{\overline{\gamma}_1\overline{\gamma}_2}{\overline{\gamma}_1 + \overline{\gamma}_2}$$

The $n$th moment can be derived using [51, eq. (3.326.2)]:

$$\mathbb{E}\left[\widetilde{\gamma}_{1(k)}^n\right]$$
$$= \frac{k}{\overline{\gamma}_1} \binom{N}{k} \sum_{m=0}^{k-1} \frac{\binom{k-1}{m}(-1)^m n!}{1 + \frac{\overline{\gamma}_2}{\overline{\gamma}}(N-k+m)}$$
$$\times \left[\overline{\gamma}_1^{n+1} + \overline{\gamma}_2\left(\frac{\rho_1\overline{\gamma} + (1-\rho_1)(N-k+m+1)\overline{\gamma}_1}{N-k+m+1}\right)^n\right] \quad (22)$$

### F. End-to-End SNDR: Fixed Gain Relaying

The relaying gain of the FG scheme is given by:

$$G \triangleq \triangleq \sqrt{\frac{\sigma^2}{\mathbb{E}\left[|\widetilde{h}_{1(k)}(t)|^2\right]P_1 + \sigma_0^2}} \quad (23)$$

where $P_1$ is the average transmitted power from $S$ and $\sigma_0^2$ is the noise variance.

The end-to-end SNDR of the FG relaying can be expressed as follows:

$$\gamma_{\text{ni}}^{\text{FG}} = \frac{\widetilde{\gamma}_{1(k)}\widetilde{\gamma}_{2(k)}}{\zeta\widetilde{\gamma}_{2(k)} + \mathbb{E}\left[\widetilde{\gamma}_{1(k)}\right] + \zeta} \quad (24)$$

where $\zeta$ is defined by:

$$\zeta = 1 + \frac{\sigma_\tau^2}{\delta^2 G^2 \sigma_0^2} \quad (25)$$

For ideal relays ($\zeta = 1$), the end-to-end SNR can be written as follows:

$$\gamma_{\text{id}}^{\text{FG}} = \frac{\widetilde{\gamma}_{1(k)}\widetilde{\gamma}_{2(k)}}{\widetilde{\gamma}_{2(k)} + \mathbb{E}\left[\widetilde{\gamma}_{1(k)}\right] + 1} \quad (26)$$

### G. End-to-End SNDR: Variable Gain Relaying I

In this relaying scheme, the relays compute the gain using the CSI of the channel $S$-$R_k$. The relays already know the CSI information since it was measured during the relay selection. However, this CSI information is not updated and it will be used to calculate the signal amplification gain which can be written as follows:

$$G \triangleq \sqrt{\frac{\sigma^2}{|\widetilde{h}_{1(k)}(t - T_d)|^2 P_1 + \sigma_0^2}} \quad (27)$$

The end-to-end SNDR is given by:

$$\gamma_{\text{ni}}^{\text{VGI}} = \frac{\widetilde{\gamma}_{1(k)}\widetilde{\gamma}_{2(k)}}{\zeta\widetilde{\gamma}_{2(k)} + \gamma_{1(k)} + \zeta} \quad (28)$$

### H. End-to-End SNDR: Variable Gain Relaying II

This relaying scheme states that unlike the VGI, the relays computes the amplification gain using the current estimated CSI rather than the outdated one. Although this scheme appears to be more realistic and sophisticated, it is very complex for implementation compared to the first version of VG since the two CSIs $h$ and $\widetilde{h}$ are required to be estimated by the control unit. The estimation of the CSI $\widetilde{h}$ is achieved by the superimposed pilots used during the feedback exchange between the various nodes of the system.

The amplification gain can be obtained by:

$$G \triangleq \sqrt{\frac{\sigma^2}{|\widetilde{h}_{1(k)}(t)|^2 P_1 + \sigma_0^2}} \quad (29)$$

In this case, the end-to-end SNDR can be derived as follows:

$$\gamma_{\text{ni}}^{\text{VGII}} = \frac{\widetilde{\gamma}_{1(k)}\widetilde{\gamma}_{2(k)}}{\zeta\widetilde{\gamma}_{2(k)} + \widetilde{\gamma}_{1(k)} + \zeta} \quad (30)$$

## III. Outage Probability Analysis

The outage probability is the probability that the overall SNDR falls below a given threshold $\gamma_{\text{th}}$ of acceptable transmission quality. It can be defined as:

$$P_{\text{out}}(\gamma_{\text{th}}) \triangleq \Pr[\gamma < \gamma_{\text{th}}] \quad (31)$$

where $\gamma$ is the effective overall SNDR and $\Pr(\cdot)$ is the probability notation.

### A. Fixed Gain Relaying

After substituting the expression of the effective SNDR (24) in Eq. (31) and applying the following identity [51, eq. (3.324.1)], the outage expression is given by:

$$P_{\text{out}}(\gamma_{\text{th}}) = 1 - \frac{2k^2}{\overline{\gamma}_1}\binom{N}{k}^2 \sum_{m=0}^{k-1}\sum_{n=0}^{k-1}\sum_{i=1}^{2}\sum_{j=1}^{2} S_m P_n T_{m,i} Q_{n,j}$$
$$\times \sqrt{\frac{U_{m,i}c\gamma_{\text{th}}}{R_{n,j}}} \exp\left(-\frac{R_{n,j}\zeta\gamma_{\text{th}}}{\overline{\gamma}_1}\right) K_1\left(\frac{2}{\overline{\gamma}_1}\sqrt{U_{m,i}R_{n,j}c\gamma_{\text{th}}}\right) \quad (32)$$

where $K_\nu(\cdot)$ is the modified Bessel function of the second kind of order $\nu$ and the parameter $c$ is given by:

$$c = \mathbb{E}\left[\overline{\gamma}_{1(k)}\right] + \zeta$$

To get a more accurate insight on the system, we derive an analytical expression of the outage probability at high SNR regime which is given by Eq. (33).

$$P_{\text{out}}^{\infty}(\gamma_{\text{th}}) \underset{\overline{\gamma}_1, \overline{\gamma}_2 \gg 1}{\cong} \frac{k^2 \gamma_{\text{th}}}{\overline{\gamma}_1} \binom{N}{k}^2 \sum_{m=0}^{k-1} \sum_{n=0}^{k-1} \sum_{i=1}^{2} \sum_{j=1}^{2} S_m P_n T_{m,i} Q_{n,j}$$
$$\times \left[\frac{T_{m,i} c}{\overline{\gamma}_1} \log\left(\frac{\overline{\gamma}_1}{R_{n,j}}\right) + \exp\left(-\frac{U_{m,i} c}{\overline{\gamma}_1}\right) + \frac{U_{m,i} c}{\overline{\gamma}_1}\right.$$
$$\left. \times \left\{1 - \gamma_{\text{e}} + \text{Ei}\left(-\frac{U_{m,i} c}{\overline{\gamma}_1}\right) - \log\left(\frac{U_{m,i} c}{\overline{\gamma}_1}\right)\right\}\right] \tag{33}$$

where $\gamma_{\text{e}}$ is the Euler-Mascheroni constant.

*Proof:* The proof of Eq. (33) is provided in appendix A. ∎

### B. Variable Gain Relaying I

In this case, we should substitute the expression of the effective SNDR (28) in Eq. (31). Since the derivation of a closed-form of the outage performance of VGI is very complex, an approximation is provided by Eq. (34).

$$P_{\text{out}}(\gamma_{\text{th}}) \cong 1 - k^2 \binom{N}{k}^2 \sum_{m=0}^{k-1} \sum_{n=0}^{k-1} \sum_{i=1}^{2} \sum_{j=1}^{2} S_m P_n T_{m,i} Q_{n,j}$$
$$\times \exp\left[-\frac{\gamma_{\text{th}}}{(1-\rho_1)\overline{\gamma}_1}\left(\frac{\rho_1(U_{m,i} - R_{n,j}\zeta)}{(1-\rho_1)R_{n,j}^2} + \zeta\right)\right]$$
$$\times \left[\left(R_{n,j}\left(1 - \frac{\rho_1}{(1-\rho_1)R_{n,j}}\right)\right)^{-1} + \frac{\gamma_{\text{th}} U_{m,i}}{(1-\rho_1)\overline{\gamma}_1 R_{n,j}^2}\right.$$
$$\left. + \log\left(\frac{1}{(1-\rho_1)\overline{\gamma}_1} - \frac{\rho_1}{(1-\rho_1)^2 \overline{\gamma}_1 R_{n,j}}\right)\right] \tag{34}$$

*Proof:* The derivation of Eq. (34) is detailed in appendix B. ∎

### C. Variable Gain Relaying II

After replacing the end-to-end SNDR (30) in Eq. (31) and after applying the identity [51, eq. (3.324.1)], the outage probability can be finally expressed as follows:

$$P_{\text{out}}(\gamma_{\text{th}}) = 1 - \frac{2k^2}{\overline{\gamma}_1} \binom{N}{k}^2 \sum_{m=0}^{k-1} \sum_{n=0}^{k-1} \sum_{i=1}^{2} \sum_{j=1}^{2} S_m P_n T_{m,i} Q_{n,j}$$
$$\times \sqrt{\frac{U_{m,i}\zeta \gamma_{\text{th}}(1+\gamma_{\text{th}})}{R_{n,j}}} \exp\left[-\frac{\gamma_{\text{th}}}{\overline{\gamma}_1}(U_{m,i} + \zeta R_{n,j})\right]$$
$$\times K_1\left(\frac{2}{\overline{\gamma}_1}\sqrt{U_{m,i} R_{n,j}\zeta \gamma_{\text{th}}(1+\gamma_{\text{th}})}\right) \tag{35}$$

For every value of $x$ very close to zero, we get $K_1(x) \approx \frac{1}{x}$ and $e^x \approx 1 + x$. Based on these asymptotic expressions, a simpler approximation of the outage expression of VGII at high-SNR regime is given by:.

$$P_{\text{out}}^{\infty}(\gamma_{\text{th}}) \underset{\overline{\gamma}_1, \overline{\gamma}_2 \gg 1}{\cong} \frac{k^2 \gamma_{\text{th}}}{\overline{\gamma}_1} \binom{N}{k}^2 \sum_{m=0}^{k-1} \sum_{n=0}^{k-1} \sum_{i=1}^{2} \sum_{j=1}^{2} S_m P_n T_{m,i} Q_{n,j}$$
$$\times \left(\zeta + \frac{U_{m,i}}{R_{n,j}}\right) \tag{36}$$

For ideal or linear relaying, the diversity gain can be derived from Eqs. (33, 34, 36). It can be expressed as follows:

$$G_{\text{d}} = \begin{cases} N, & \rho_1 = \rho_2 = 1 \\ 1, & \rho_1, \rho_2 < 1 \end{cases}$$

If the relays are impaired, the outage performance saturates by the impairments floor and so the diversity gain in this case is equal to zero ($G_{\text{d}} = 0$).

## IV. AVERAGE BIT ERROR RATE ANALYSIS

In this section, we address the error performance of the system for different modulation schemes and considering the three relaying modes. The average BER for various modulation formats such as BPSK, $M$-PAM, $M$-PSK and $M$-QAM is defined by:

$$\overline{P}_{\text{e}} = \alpha \mathbb{E}\left[Q(\sqrt{2\beta\gamma})\right] \tag{37}$$

where $Q(x) = \frac{1}{\sqrt{2\pi}}\int_x^{\infty} e^{-\frac{t^2}{2}} dt$ is the Gaussian Q-function and $\alpha, \beta$ are the modulation parameters. Using integration by parts, Eq. (37) can be expressed as follows:

$$\overline{P}_{\text{e}} = \frac{\alpha\sqrt{\beta}}{2\sqrt{\pi}}\int_0^{\infty} \frac{e^{-\beta\gamma}}{\sqrt{\gamma}} F_\gamma(\gamma) d\gamma \tag{38}$$

### A. Fixed Gain Relaying

To derive a closed-form of the average BER for the FG relaying scheme, we should substitute the expression of the outage probability (32) in Eq. (38). Then we must apply the identity [56, eq. (4.17.37)] to get the expression as follows:

$$\overline{P}_{\text{e}} = \frac{\alpha}{2} - \frac{\alpha k^2}{2}\sqrt{\frac{\beta\overline{\gamma}_1}{\pi}}\Gamma\left(\frac{1}{2}\right)\Gamma\left(\frac{3}{2}\right)\binom{N}{k}^2$$
$$\times \sum_{m=0}^{k-1} \sum_{n=0}^{k-1} \sum_{i=1}^{2} \sum_{j=1}^{2} \frac{S_m P_n T_{m,i} Q_{n,j}}{R_{n,j}\sqrt{2R_{n,j}\zeta + \beta\overline{\gamma}_1}}$$
$$\times \exp\left(\frac{U_{m,i} R_{n,j}\zeta}{\overline{\gamma}_1(\beta\overline{\gamma}_1 + 2\zeta R_{n,j})}\right) W_{-\frac{1}{2},\frac{1}{2}}$$
$$\times \left(\frac{2U_{m,i} R_{n,j}}{\overline{\gamma}_1(\beta\overline{\gamma}_1 + 2\zeta R_{n,j})}\right) \tag{39}$$

where $W_{p,q}(\cdot)$ is the Whittaker function.

Now, we should substitute Eq. (33) in (38). After applying the identity [57, eq. (2.3.3.1)], the high SNR approximation of the average BER of FG relaying can be expressed as follows:

$$
\overline{P_e}^\infty \underset{\overline{\gamma_1}, \overline{\gamma_2} \gg 1}{\cong} \frac{\alpha k^2}{2\overline{\gamma_1}} \binom{N}{k}^2 \sum_{m=0}^{2} \sum_{n=0}^{2} \sum_{i=1}^{2} \sum_{j=1}^{2} S_m P_n T_{m,i} Q_{n,j}
$$
$$
\times \left[ \frac{U_{m,i}c}{\overline{\gamma_1}} \log\left(\frac{\overline{\gamma_1}}{R_{n,j}}\right) + \exp\left(-\frac{U_{m,i}c}{\overline{\gamma_1}}\right) + \frac{U_{m,i}c}{\overline{\gamma_1}} \right.
$$
$$
\left. \times \left\{ 1 - \gamma_e + \mathrm{Ei}\left(-\frac{U_{m,i}c}{\overline{\gamma_1}}\right) - \log\left(\frac{U_{m,i}c}{\overline{\gamma_1}}\right) \right\} \right]
$$
(40)

### B. Variable Gain Relaying I

After substituting the expression (34) in Eq. (38) and applying the identity [57, eq. (2.3.3.1)], the approximation of the average BER can be derived as follows:

$$
\overline{P_e} \cong \frac{\alpha}{2} - \frac{\alpha k^2}{2} \binom{N}{k}^2 \sum_{n=0}^{2} \sum_{i=1}^{2} \sum_{j=1}^{2} S_m P_n T_{m,i} Q_{n,j}
$$
$$
\times \sqrt{\frac{\beta}{2\mu + \beta}} \left[ \eta + \frac{\nu}{2\mu + \beta} \right]
$$
(41)

where $\eta$, $\mu$ and $\nu$ are given by:

$$
\eta = \frac{1}{R_{n,j} - \frac{\rho_1}{1 - \rho_1}}
$$
$$
\mu = \frac{1}{(1-\rho_1)\overline{\gamma_1}} \left( \frac{\rho_1(U_{m,i} - R_{n,j}\zeta)}{(1-\rho_1)R_{n,j}^2} + \zeta \right)
$$
$$
\nu = \frac{U_{m,i}}{(1-\rho_1)\overline{\gamma_1}R_{n,j}^2} \log\left( \frac{1}{(1-\rho_1)^2} - \frac{\rho_1}{(1-\rho_1)^2\overline{\gamma_1}R_{n,j}} \right)
$$

### C. Variable Gain Relaying II

Since the derivation of a closed-form of the average BER is complex, we should consider a simpler form. After some mathematical manipulation, the analytical approximation is given by Eq. (42).

$$
\overline{P_e} = \frac{\alpha}{2} - \frac{\alpha k^2 \sqrt{2}}{\beta \overline{\gamma_1}} \Gamma\left(\frac{1}{2}\right) \Gamma\left(\frac{5}{2}\right) \binom{N}{k}^2
$$
$$
\times \sum_{m=0}^{k-1} \sum_{n=0}^{k-1} \sum_{i=1}^{2} \sum_{j=1}^{2} S_m P_n T_{m,i} Q_{n,j} \sqrt{\frac{\zeta U_{m,i}}{R_{n,j}}} \frac{\varrho}{(\omega + \varrho)^{\frac{5}{2}}}
$$
$$
\times {}_2F_1\left(\frac{5}{2}, \frac{3}{2}, 2, \frac{\omega - \varrho}{\omega + \varrho}\right)
$$
(42)

where ${}_pF_q(\mathrm{a,b,z})$ is the hypergeometric function.

*Proof:* The proof is detailed in appendix C. ∎

After substituting the expression (36) in Eq. (38) and applying the identity [57, eq. (2.3.3.1)], the asymptotic high SNR of the BER is given by:

$$
\overline{P_e}^\infty \underset{\overline{\gamma_1}, \overline{\gamma_2} \gg 1}{\cong} \frac{\alpha k^2}{2\beta \overline{\gamma_1}} \binom{N}{k}^2 \sum_{m=0}^{k-1} \sum_{n=0}^{k-1} \sum_{i=1}^{2} \sum_{j=1}^{2} S_m P_n T_{m,i} Q_{n,j}
$$
$$
\times \left(\zeta + \frac{U_{m,i}}{R_{n,j}}\right)
$$
(43)

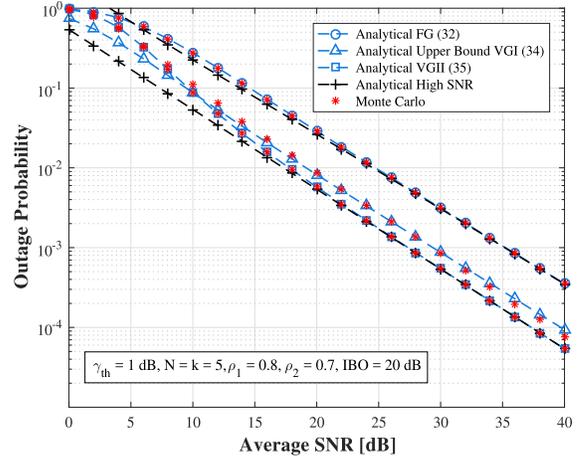

Fig. 3. Outage probabilities of FG, VGI and VGII relaying under the SEL hardware impairment.

## V. ERGODIC CAPACITY ANALYSIS

The channel capacity, expressed in (bit/s/Hz), is defined as the maximum error-free data rate transmitted by the system. It can be written as follows:

$$
\overline{C} = \frac{1}{2}\mathbb{E}\left[\log_2(1 + \gamma)\right]
$$
(44)

Since the transmission is achieved in two steps, the system capacity is multiplied by the factor $\frac{1}{2}$. After some mathematical manipulation, the egodic capacity can be expressed as follows:

$$
\overline{C} = \frac{1}{2\log(2)} \int_0^\infty \frac{\overline{F_\gamma}(\gamma)}{\gamma + 1} d\gamma
$$
(45)

where $\gamma$ is the end-to-end SNDR and $\overline{F_\gamma}$ is the complementary cumulative distribution function (CCDF) of $\gamma$.

Since the non-linear distortion deteriorates the system performance, an indesirable ceiling is created by the impairments which limits the achievable rate of the system. The ceiling expression is given by [34, eq. (37)]:

$$
\overline{C}^* = \frac{1}{2}\log_2\left(1 + \frac{1}{\frac{\varepsilon}{\delta^2} - 1}\right)
$$
(46)

where $\varepsilon$ is the clipping factor of the hardware impairments.

### A. Fixed Gain Relaying

After replacing the CCDF of the SNDR (24) in (46) and applying some mathematical manipulation, the closed-form of the ergodic capacity is derived in term of bivariate Meijer G-function as follows:

$$
\overline{C} = \frac{k^2\binom{N}{k}^2}{2\log(2)\overline{\gamma_1}} \sum_{m=0}^{k-1} \sum_{n=0}^{k-1} \sum_{i=1}^{2} \sum_{j=1}^{2} S_m P_n T_{m,i} Q_{n,j} \sqrt{\frac{U_{m,i}c}{R_{n,j}}}
$$
$$
\times G_{1,1:0:1:0:2}^{1,1:1,0:2,0}\left(\begin{matrix} -\frac{1}{2} \\ -\frac{1}{2} \end{matrix} \middle| \begin{matrix} - \\ 0 \end{matrix} \middle| \begin{matrix} - \\ \frac{1}{2}, -\frac{1}{2} \end{matrix} \middle| \frac{R_{n,j}\zeta}{\overline{\gamma_1}}, \frac{U_{m,i}R_{n,j}c}{\overline{\gamma_1}^2}\right)
$$
(47)

*Proof:* The derivation steps of Eq. (47) are given in appendix D. ∎

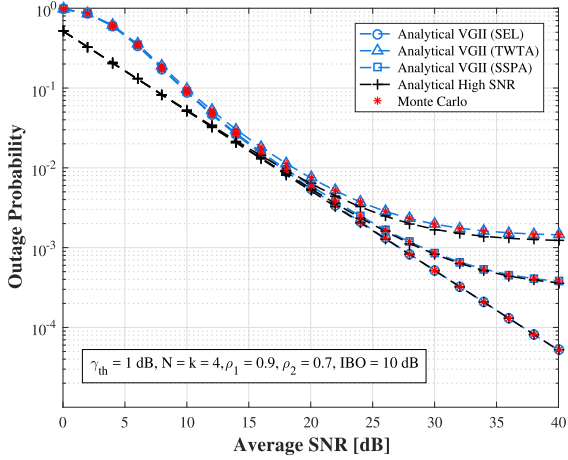

Fig. 4. Outage probability of VGII relaying under the SEL, TWTA and SSPA impairments.

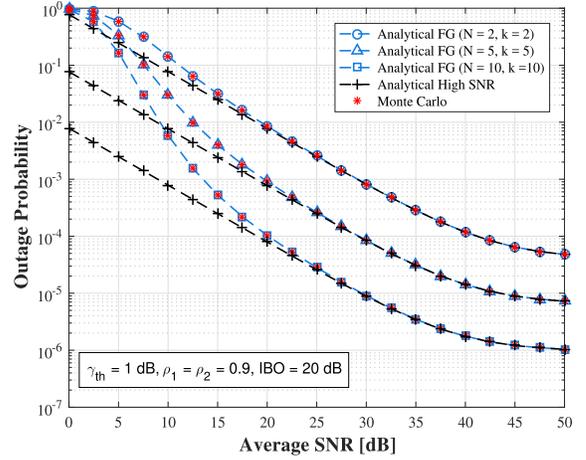

Fig. 5. Outage probability of FG relaying for various number of relays.

## B. Variable Gain Relaying I

In this case, we should replace the expression of the CCDF of (28) in Eq. (45). After referring to the identity [51, eq. (3.353.5)], the approximation of the capacity is derived as follows:

$$\overline{C}^{\text{ub}} \cong \frac{k^2}{2 \log(2)} \binom{N}{k}^2 \sum_{m=0}^{k-1} \sum_{n=0}^{k-1} \sum_{i=1}^{2} \sum_{j=1}^{2} S_m P_n T_{m,i} Q_{n,j}$$
$$\times \left[ \frac{\nu}{\mu} + e^\mu \text{Ei}(-\mu)(\nu - \eta) \right] \quad (48)$$

## C. Variable Gain Relaying II

Since the integral (45) is not solvable for the case of VGII, we derive a very tight upper bound in term of bivariate Fox H-function.

$$\overline{C}^{\text{ub}} \cong \frac{k^2 \overline{\gamma}_1 \binom{N}{k}^2}{4 \log(2)} \sum_{m=0}^{k-1} \sum_{n=0}^{k-1} \sum_{i=1}^{2} \sum_{j=1}^{2} \frac{S_m P_n T_{m,i} Q_{n,j}}{(U_{m,i} + \zeta R_{n,j})^2} \sqrt{\frac{U_{m,i} \zeta}{R_{n,j}}}$$
$$\times H_{1,0:1:1:0,2}^{0,1:1,1:2,0} \left( \begin{matrix} (-1; 1, 1) \\ - \end{matrix} \Big| \begin{matrix} (0, 1) \\ (0, 1) \end{matrix} \Big| \begin{matrix} - \\ (\frac{1}{2}, \frac{1}{2}), (-\frac{1}{2}, \frac{1}{2}) \end{matrix} \Big| \tau_1, \tau_2 \right) \quad (49)$$

where $\tau_1$, $\tau_2$ are defined by:

$$\tau_1 = \frac{\overline{\gamma}_1}{U_{m,i} + \zeta R_{n,j}}; \quad \tau_2 = \frac{\sqrt{\zeta U_{m,i} R_{n,j}}}{U_{m,i} + \zeta R_{n,j}}$$

*Proof:* The proof is given in appendix E.                                    ∎

## VI. NUMERICAL RESULTS AND DISCUSSION

In this section, we present the analytical and simulation results illustrating the effects of the hardware impairments, the relaying schemes, the number of the relays, the rank of the selected relay and the outdated CSI on the system. The performance metrics used to quantify the robustness and the resiliency of the system, are the outage probability, the average bit error rate and the ergodic capacity. The analytical results are confirmed by Monte Carlo simulation considering $10^9$ iterations.

Fig. 3 shows the variations of the outage probability of FG, VGI and VGII with respect to the average SNR. As expected, it is clear that the variable gain relaying outperforms the FG scheme. Regarding the variable gain protocol, the system performs better when using the second version compared to the first one. In fact, the main difference between the two versions is the CSI used for the relaying amplification. Given that the second version employs the perfect CSI retrieved by the pilot training technique, the amplification in the first version is based on the outdated CSI. As a result, the CSI used for the amplification makes the second version of the variable gain relaying more efficient than the first one.

Fig. 4 presents the dependence of the outage performance of VGII relaying against the average SNR under the different models of impairment. For low SNR, the system response to the impairment is acceptable as the three impairments' models have the same impact. As the average SNR increases above 20 dB, the system responses to the various hardware impairments significantly differ from each other. We note that in the high SNR region, the impairments effect becomes more severe particularly for the TWTA and SSPA. As the average SNR exceeds 25 dB, an irreducible outage floor is created which inhibits the performance from converging to zero. Graphically, we note that the system saturates at 0.002 and 0.0003, respectively, for TWTA and SSPA. Consequently, the TWTA has the most detrimental effect on the system. For the SEL impairment model, the system still operates in acceptable conditions and there is no significant impact on the system performance especially the non-creation of the outage floor unlike SSPA and TWTA at least below 40 dB.

Fig. 5 shows the variations of the outage probability of FG relaying against the average SNR under the SSPA impairment and for various number of relays. For low SNR and below 10 dB, the number of relays has no remarkable impact on the outage probability. However, as the SNR grows largely, the performance significantly deviates from each other. In fact, the system operates better as the number of relays increases. To achieve an outage probability equal to $10^{-3}$, the system requires the following average SNRs 20 dB, 27 dB and 35 dB, respectively, for N = 10, 5 and 2 relays. Thereby, the main contribution of the number of relays is useful to reduce the power consumption of the system. This main advantage is explained by the fact that for a higher number of relays, there is a higher probability to select a better channel/relay. However, as the average SNR increases, the impairments effect becomes more severe as the outage probability saturates by the irreducible floor created by the impairments. Even the number of relays play no significant role in this situation. Therefore, the number of relays introduces limited improvements at low SNR, however, it does not contribute in anyway as the impairments become severe at high SNR.

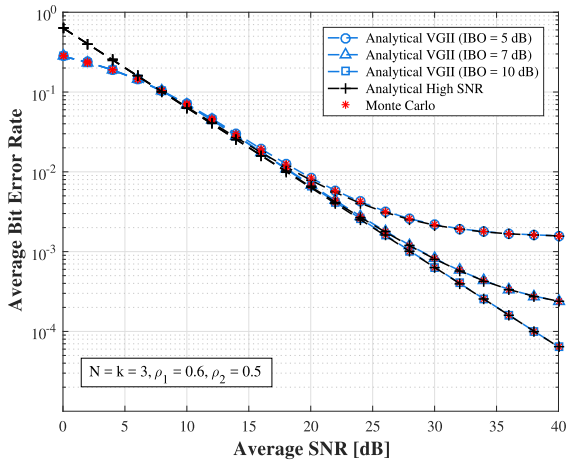

Fig. 6. Average Bit Error Rate of VGII relaying for various IBO levels.

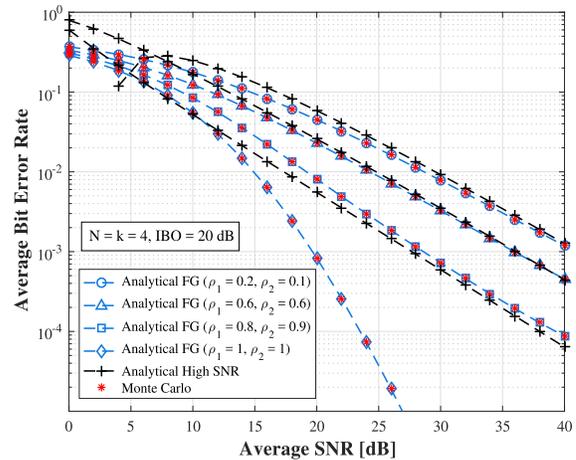

Fig. 7. Average Bit Error Rate of FG relaying for various correlation values under the SSPA impairment.

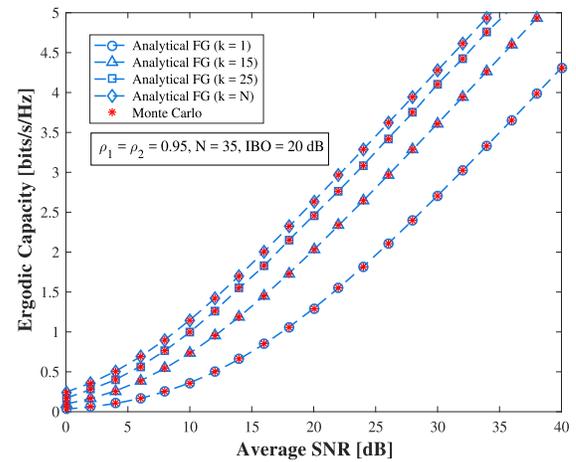

Fig. 8. Ergodic capacity of FG relaying for various ranks of the selected relay and low correlation.

Fig. 6 illustrates the variations of the BER of VGII relaying under the SEL impairment and for different values of the IBO. For low SNR below 20 dB, the IBO factor has no observable impact on the system, i.e, the BER is the same regardless of the IBO values. However, when the average SNR overtakes 25 dB, the IBO factor gets more involved. In fact, as the IBO value increases, the system performs better. For lower value of IBO = 5 dB, the BER is limited by a floor created at higher value of the SNR. Considering a large value of IBO = 10 dB, the system performance improves and the BER floors are mitigated. Technically, increasing the IBO value comes directly from increasing the input saturation level $A_{sat}$. We already showed that the saturation's amplifier is relieved as the input saturation level increases. For a lower value of $A_{sat}$, i.e, lower value of IBO, the system becomes more saturated by the impairment's distortion. Consequently, the relation between the input saturation level and the IBO thoroughly explains the impact of higher values of IBO on the system performance.

Fig. 7 illustrates the variations of the average BER of FG relaying under the SSPA impairment and for different values of the correlation coefficients $\rho_1$ and $\rho_2$. We note that the system performs better as the correlation coefficients increase. In fact, both the arrangement and the selection of the relay are based on the CSI monitored by the control unit. As the correlation coefficients grow, the CSI estimation becomes more accurate and so the relay selection will be based on error-free CSI estimation. Furthermore, when we achieve a full correlation between the CSIs ($\rho_1, \rho_2 \approx 1$), the performance improves further particularly when the relay of the last rank is selected. However, when the correlation coefficients decrease, i.e, the CSIs become more uncorrelated, the relay selection will be based on a completely outdated CSI. In this case, even when we select the relay of the last rank $N$, the performance gets worse since the selection of the best relay becomes uncertain and there is no relation between the received CSI and the rank of the selected relay.

The same results given by Fig. 7 are confirmed by other approaches in figures 8 and 9 which present the variations of the channel capacity for different values of $k$ and for high and low correlation coefficients, respectively. Unlike the configuration assumed in Fig. 7, the correlation coefficients ($\rho_1, \rho_2$) are fixed to a high value (0.95) and the rank $k$ is varied. We note that the capacity performance significantly enhances when the rank $k$ increases. Given that we assumed the opportunistic protocol for the relay selection, we stated that the control unit arranges the CSIs in an increasing order. Thereby, as the rank of the selected relay becomes closer to the rank of the best relay (rank = $N$), the system performs better. In this case, the efficiency of the channel/relay is related to the rank given that the correlation must be high. However, the results of Fig. 9 are absolutely the opposite for the configuration adopted in Fig. 8. We clearly see that the system performs worse as the rank $k$ becomes higher. In fact, this result is expected since the CSIs are completely uncorrelated (low correlation 0.009) and so the rank $k$ has nothing to do with the channel/relay efficiency.

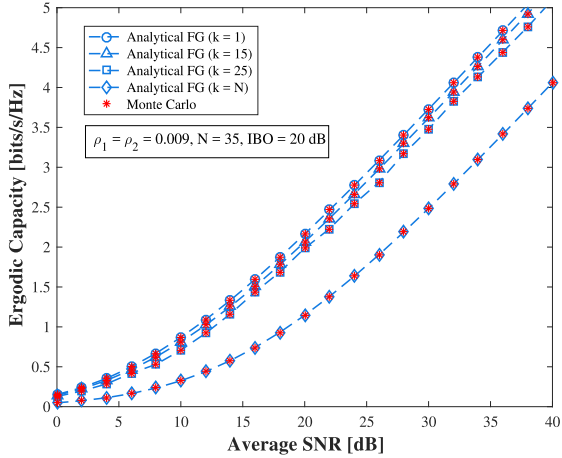

Fig. 9.　Ergodic capacity of FG relaying for various ranks of the selected relay and high correlation.

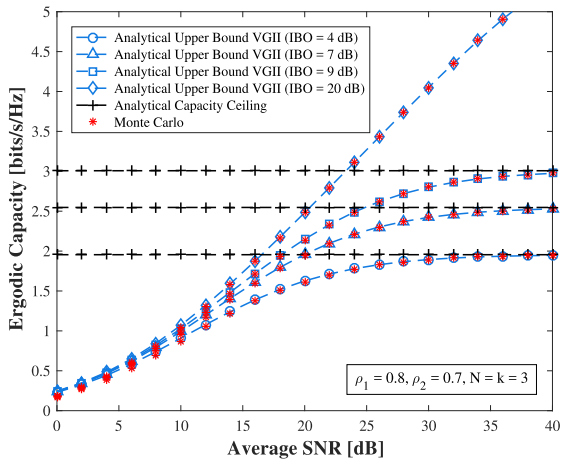

Fig. 10.　Ergodic capacity of VGII relaying for various IBO levels under the TWTA impairment.

The effect of IBO is illustrated by Fig. 10 which presents the variations of the ergodic capacity for different values of IBO. As we concluded about the effect of IBO on the BER performance in Fig. 6, the impact of IBO is more notable on the capacity performance at high SNR. As the IBO decreases, the channel capacity saturates more especially for IBO = 4 dB and the maximum rate is around 2 bits/s/Hz. However, the level of saturation vanishes for a higher value of IBO equal to 20 dB. For low SNR, the effect of IBO is negligeable and the system operates efficiently. This result is graphically shown by the small difference between the capacities for different values of IBO, especially for an average SNR range less than 15 dB. As the average SNR increases, the IBO essentially contributes to improve the extent of the achievable rate.

## VII. Conclusion

In this work, we present a system with multiple relays operating at various relaying schemes FG, VGI and VGII.

We assume the opportunistic relay selection to choose a single relay to forward the signal. Moreover, we introduce three models of the hardware impairments SEL, TWTA and SSPA that affect the relays during the power amplification. We quantify the impacts of these imperfections on the system performance in terms of the outage probability, the average BER and the ergodic capacity. We also investigate the effects of the IBO, the number of relays, the rank of the selected relay and the correlation coefficients on the system. We conclude that the impairments have deleterious impacts on the system as the average SNR increases and particularly the TWTA impairment model has the most detrimental effects on the system compared to SSPA and SEL. We also demonstrate that as the number of relays increases, the performance substantially improves mainly the power consumption significantly decreases. Furthermore, we show that the system performs better when selecting the relay with the highest rank simultaneously coupled with higher values of the correlation coefficients. In addition, we prove that the capacity saturates quickly at high SNR when the IBO level is low and grows up infinitely as the IBO takes higher values.

## Appendix A
## High SNR Approximation - Fixed Gain Relaying

The end-to-end SNDR $\gamma_{\mathrm{ni}}^{\mathrm{FG}}$ is upper bounded by $\gamma_{\mathrm{u}}$ which is given by:

$$\gamma_{\mathrm{u}} = \min\left\{\widetilde{\gamma}_{1(k)}, \frac{\widetilde{\gamma}_{1(k)}\widetilde{\gamma}_{2(k)}}{C}\right\} \qquad (50)$$

The complementary CDF (CCDF) of $\gamma_{\mathrm{u}}$ can be written as follows:

$$\overline{F}_{\mathrm{u}}(\gamma_{\mathrm{th}}) = \Pr(\gamma_{\mathrm{u}} > \gamma_{\mathrm{th}}) = \Pr\left(\frac{\widetilde{\gamma}_{1(k)}\widetilde{\gamma}_{2(k)}}{C} > \gamma_{\mathrm{th}} \cap \widetilde{\gamma}_{1(k)} > \gamma_{\mathrm{th}}\right)$$

$$= \int_{\gamma_{\mathrm{th}}}^{\infty} f_{\widetilde{\gamma}_{1(k)}}(x)\overline{F}_{\widetilde{\gamma}_{2(k)}}\left(\frac{C\gamma_{\mathrm{th}}}{x}\right)dx \qquad (51)$$

The high SNR approximation is nothing but the CDF of $\gamma_{\mathrm{u}}$, given by $F_{\mathrm{u}}(\gamma_{\mathrm{th}}) = 1 - \overline{F}_{\mathrm{u}}(\gamma_{\mathrm{th}})$.

After developing this expression, the CCDF of $\gamma_{\mathrm{u}}$ can be written as the summation of four integrals where each of them has the following general form:

$$I = \int_{\gamma_{\mathrm{th}}}^{\infty} e^{-\left(\frac{a}{\overline{\gamma}_2 x} + \frac{bx}{\overline{\gamma}_1}\right)} = \sum_{r=0}^{\infty}\left(-\frac{a}{\overline{\gamma}_2}\right)^r \frac{1}{r!}\int_{\gamma_{\mathrm{th}}}^{\infty}\frac{e^{-\frac{bx}{\overline{\gamma}_1}}}{x^r}dx$$

$$= \sum_{r=0}^{\infty}\left(-\frac{a}{\overline{\gamma}_2\gamma_{\mathrm{th}}}\right)^r \frac{1}{r!}e^{-\frac{b\gamma_{\mathrm{th}}}{\overline{\gamma}_1}}\int_{\gamma_{\mathrm{th}}}^{\infty}\frac{e^{-\frac{bx}{\overline{\gamma}_1}}}{\left(\frac{x}{\gamma_{\mathrm{th}}}+1\right)}dx \qquad (52)$$

As the average SNRs $\overline{\gamma}_1$ and $\overline{\gamma}_2$ grow largely, we can approximate the expression of the integral $I$ by using [58, eq. (25)].

$$I \approx e^{-\frac{b\gamma_{\mathrm{th}}}{\overline{\gamma}_1}}\left(\frac{\overline{\gamma}_1}{b} + \frac{a}{\overline{\gamma}_2}\log\left(\frac{b}{\overline{\gamma}_1}\right) + \sum_{r=2}^{\infty}\left(-\frac{a}{\overline{\gamma}_2}\right)^r\frac{\gamma_{\mathrm{th}}^{1-r}}{r!(r-1)}\right) \qquad (53)$$

After that, we apply the following identity for every $x \neq 0$:

$$\text{Ei}(x) = \gamma_e + \log|x| + \sum_{r=1}^{\infty} \frac{x^r}{r \, r!} \qquad (54)$$

After some mathematical manipulation, we finally derive the asymptotic high SNR.

## Appendix B
## Outage Probability Derivation - Variable Gain Relaying I

It is complex to derive a closed-form of the outage probability of VGI. In this case, we have to derive an approximation of the end-to-end SNDR $\gamma_{\text{ni}}^{\text{VGI}}$.

$$\gamma_{\text{ni}}^{\text{VGI}} \approx \frac{\widetilde{\gamma}_{1(k)} \widetilde{\gamma}_{2(k)}}{\gamma_{1(k)} + \zeta \, \widetilde{\gamma}_{2(k)}} \qquad (55)$$

The approximate outage probability can be written as follows:

$$P_{\text{out}}(\gamma_{\text{th}}) \approx 1 - \int_0^{\infty} \int_0^{\infty} f_{\widetilde{\gamma}_{1(k)}, \gamma_{1(k)}}(x + \zeta \, \gamma_{\text{th}}, y) \\ \times \left(1 - F_{\widetilde{\gamma}_{2(k)}}\left(\frac{\gamma_{\text{th}} \, y}{x}\right)\right) dy \, dx \qquad (56)$$

where $f_{\widetilde{\gamma}_{1(k)}, \gamma_{1(k)}}(x, y)$ is the joint PDF of the two random variables $\gamma_{1(k)}$ and $\widetilde{\gamma}_{1(k)}$ given by [22, eq. (48)]. After substituting the expression of the joint PDF in Eq. (56), the approximation can be written as the summation of integrals taking the general form as follows:

$$P_{\text{out}}(\gamma_{\text{th}}) \approx 1 - \frac{k^2 \binom{N}{k}^2}{(1-\rho_1) \overline{\gamma}_1^2} \sum_{m=0}^{k-1} \sum_{n=0}^{k-1} \frac{(-1)^{n+m} \binom{k-1}{m}}{1 + \frac{\overline{\gamma}_1}{\overline{\gamma}}(N-k+m)} \\ \times \frac{\binom{k-1}{n}}{1 + \frac{\overline{\gamma}_2}{\overline{\gamma}}(N-k+n)} \sum_r \alpha_r I_r \qquad (57)$$

where $I_r$ is given by:

$$I_r = \int_0^{\infty} \int_0^{\infty} e^{-\frac{x + \zeta \gamma_{\text{th}}}{(1-\rho_1) \overline{\gamma}_1}} e^{-\frac{y}{\overline{\gamma}_1}\left(a_r + \frac{b_r \gamma_{\text{th}}}{x}\right)} \\ \times I_0\left(\frac{2\sqrt{\rho_1 (x + \zeta \gamma_{\text{th}}) y}}{(1-\rho_1) \overline{\gamma}_1}\right) dy \, dx \qquad (58)$$

We can simplify further the expression of $I_r$, we get:

$$I_r = \int_0^{\infty} e^{-\frac{x + \zeta \gamma_{\text{th}}}{(1-\rho_1) \overline{\gamma}_1}\left(1 - \frac{\rho_1}{(1-\rho_1)\left(a_r + \frac{b_r \gamma_{\text{th}}}{x}\right)}\right)} \left(a_r + \frac{b_r \gamma_{\text{th}}}{x}\right)^{-1} dx \qquad (59)$$

Since a closed-form of the integral $I_r$ does not exist, we should apply a partial fraction expansion on the argument of the exponential function to get a simpler form of $I_r$.

$$\frac{x + \zeta \gamma_{\text{th}}}{(1-\rho_1) \overline{\gamma}_1} \left(1 - \frac{\rho_1}{(1-\rho_1)\left(a_r + \frac{b_r \gamma_{\text{th}}}{x}\right)}\right) \\ = x \left(\frac{1}{(1-\rho_1) \overline{\gamma}_1} - \frac{\rho_1}{a_r \overline{\gamma}_1 (1-\rho_1)^2}\right)$$

$$+ \frac{\gamma_{\text{th}}}{(1-\rho_1) \overline{\gamma}_1} \left(\frac{\rho_1 (b_r - a_r \zeta)}{(1-\rho_1) a_r^2} + \zeta\right) \\ + \frac{\gamma_{\text{th}}^2 \rho_1 b_r (a_r \zeta - b_r)}{a_r^2 \overline{\gamma}_1 (1-\rho_1)^2 (b_r \gamma_{\text{th}} + a_r x)} \qquad (60)$$

Applying the Maclaurin series over the following term $e^{-\frac{\gamma_{\text{th}}^2 \rho_1 b_r (a_r \zeta - b_r)}{a_r^2 \overline{\gamma}_1 (1-\rho_1)^2 (b_r \gamma_{\text{th}} + a_r x)}}$ and ignoring higher order of $\frac{\gamma_{\text{th}}}{\overline{\gamma}_1}$ to get the following approximation

$$I_r = e^{-\frac{\gamma_{\text{th}}}{(1-\rho_1) \overline{\gamma}_1}\left(\frac{\rho_1 (b_r - a_r \zeta)}{(1-\rho_1) a_r^2} + \zeta\right)} \int_0^{\infty} e^{-x\left(\frac{1}{(1-\rho_1) \overline{\gamma}_1} - \frac{\rho_1}{a_r \overline{\gamma}_1 (1-\rho_1)^2}\right)} \\ \times \left(\frac{1}{a_r} - \frac{\gamma_{\text{th}} b_r}{a_r (a_r x + b_r \gamma_{\text{th}})}\right) dx + O\left(\left(\frac{\gamma_{\text{th}}}{\overline{\gamma}_1}\right)^2\right) \qquad (61)$$

Using [58, eq. (11)], the approximate form of the integral $I_r$ can be developped as follows:

$$I_r = e^{-\frac{\gamma_{\text{th}}}{(1-\rho_1) \overline{\gamma}_1}\left(\frac{\rho_1 (b_r - a_r \zeta)}{(1-\rho_1) a_r^2} + \zeta\right)} \left(\left[a_r \left(1 - \frac{\rho_1}{(1-\rho_1) a_r}\right)\right]^{-1} \\ + \frac{\gamma_{\text{th}} b_r}{(1-\rho_1) \overline{\gamma}_1 a_r^2} \log\left(\frac{1}{(1-\rho_1) \overline{\gamma}_1} - \frac{\rho_1}{(1-\rho_1)^2 \overline{\gamma}_1 a_r}\right)\right) \qquad (62)$$

After some mathematical manipulation, we derive the approximation of the outage probability of VGI relaying scheme.

## Appendix C
## Average Bit Error Rate - Variable Gain Relaying II

After substituting the expression of the outage probability given by (35) in Eq. (38), the resulting integral function is not solvable. In this case, it is practical to provide an approximation of the BER. The first step is to modify the expression of the end-to-end SNDR $\gamma_{\text{ni}}^{\text{VGII}}$ as follows:

$$\gamma_{\text{ni}}^{\text{VGII}} \approx \frac{\widetilde{\gamma}_{1(k)} \widetilde{\gamma}_{2(k)}}{\widetilde{\gamma}_{1(k)} + \zeta \, \widetilde{\gamma}_{2(k)}} \qquad (63)$$

Using the identity given by [51, eq. (6.621.3)] and after some mathematical manipulation, we finally derive the analytical expression of the approximation of VGII relaying gain.

## Appendix D
## Ergodic Capacity - Fixed Gain Relaying

To derive the expression of the system capacity, we should substitute the expression of the CCDF of (24) in Eq. (45). Then we apply the identities [55, eqs. (07.34.03.0271.01), (07.34.03.0046.01), and (03.04.26.0009.01)] to transform the fraction, exponential and Bessel functions, respectively, into Meijer G-function. Then, we refer to the identity [59, eq. (9)] to solve the integral containing three Meijer G-functions. After some mathematical manipulation, the closed-form of the ergodic capacity is derived in term of bivariate Meijer G-function.

The implementation of the bivariate Meijer G-function in Matlab can be found in [60].

## Appendix E
### Ergodic Capacity - Variable Gain Relaying II

First of all, we consider an upper bound of the end-to-end SNDR (63). Then we compute the approximate CDF by substituting the new expression of the end-to-end SNDR by applying the identity [51, eq. (3.324.1)]. After that, placing the relative upper bound of CCDF in Eq. (45) and applying the identities [55, eqs. (07.34.03.0271.01), (07.34.03.0046.01), and (03.04.26.0009.01)] to transform the fraction, exponential and Bessel functions, respectively, into Meijer G-function. Then we combine the identities [55, eqs. (07.35.26.0003.01) and (07.35.26.0004.001)] to transform the Meijer G-function to Fox H-function. Finally, using the identity [61, eq. (2.3)] to evaluate the analytical expression of the integral containing three Fox H-functions. After some mathematical manipulation, the ergodic capacity is derived in term of bivariate Fox H-function.

An efficient implementation of bivariate Fox H-function in Matlab can be found in [62].

### References

[1] C. Hoymann, W. Chen, J. Montojo, A. Golitschek, C. Koutsimanis, and X. Shen, "Relaying operation in 3GPP LTE: Challenges and solutions," *IEEE Commun. Mag.*, vol. 50, no. 2, pp. 156–162, Feb. 2012.

[2] I. Krikidis and J. Thompson, *Cooperative Communications for improved wireless network transmission: Framework for Virtual Antenna Array Applications.* Hershey, PA, USA: IGI Global Editions, 2009.

[3] R. Pabst *et al.*, "Relay-based deployment concepts for wireless and mobile broadband radio," *IEEE Wireless Commun. Mag.*, vol. 42, no. 9, pp. 80–89, Sep. 2004.

[4] Y. Li, B. Vucetic, Z. Zhou, and M. Dohler, "Distributed adaptive power allocation for wireless relay networks," in *Proc. IEEE Int. Conf. Commun.*, Jun. 2007, pp. 5235–5240.

[5] J. N. Laneman, D. N. C. Tse, and G. W. Wornell, "Cooperative diversity in wireless networks: Efficient protocols and outage behavior," *IEEE Trans. Inf. Theory*, vol. 50, no. 12, pp. 3062–3080, Dec. 2004.

[6] A. Chandra, C. Bose, and M. K. Bose, "Wireless relays for next generation broadband networks," *IEEE Potentials*, vol. 30, no. 2, pp. 39–43, Mar. 2011.

[7] Y. Yang, H. Hu, J. Xu, and G. Mao, "Relay technologies for WiMax and LTE-advanced mobile systems," *IEEE Commun. Mag.*, vol. 47, no. 10, pp. 100–105, Oct. 2009.

[8] Y. Hua, D. W. Bliss, S. Gazor, Y. Rong, and Y. Sung, "Guest editorial theories and methods for advanced wireless relays—Issue I," *IEEE J. Sel. Areas Commun.*, vol. 30, no. 8, pp. 1297–1303, Sep. 2012.

[9] M. O. Hasna and M. S. Alouini, "A performance study of dual-hop transmissions with fixed gain relays," *IEEE Trans. Wireless Commun.*, vol. 3, no. 6, pp. 1963–1968, Nov. 2004.

[10] M. O. Hasna and M. S. Alouini, "Harmonic mean and end-to-end performance of transmission systems with relays," *IEEE Trans. Commun.*, vol. 52, no. 1, pp. 130–135, Jan. 2004.

[11] N. Y. Ermolova and S. G. Haggman, "An extension of Bussgang's theory to complex-valued signals," in *Proc. 6th Nordic Signal Process. Symp. (NORSIG)*, Jun. 2004, pp. 45–48.

[12] E. Lee, J. Park, D. Han, and G. Yoon, "Performance analysis of the asymmetric dual-hop relay transmission with mixed RF/FSO links," *IEEE Photon. Technol. Lett.*, vol. 23, no. 21, pp. 1642–1644, Nov. 1, 2011.

[13] M. I. Petkovic, A. M. Cvetkovic, G. T. Djordjevic, and G. K. Karagiannidis, "Partial relay selection with outdated channel state estimation in mixed RF/FSO systems," *J. Lightw. Technol.*, vol. 33, no. 13, pp. 2860–2867, Jul. 1, 2015.

[14] E. Soleimani-Nasab and M. Uysal, "Generalized performance analysis of mixed RF/FSO cooperative systems," *IEEE Wireless Commun.*, vol. 15, no. 1, pp. 714–727, Jan. 2016.

[15] H. Samimi and M. Uysal, "End-to-end performance of mixed RF/FSO transmission systems," *IEEE/OSA J. Opt. Commun. Netw.*, vol. 5, no. 11, pp. 1139–1144, Nov. 2013.

[16] N. I. Miridakis, M. Matthaiou, and G. K. Karagiannidis, "Multiuser relaying over mixed RF/FSO links," *IEEE Trans. Commun.*, vol. 62, no. 5, pp. 1634–1645, May 2014.

[17] N. Varshney and A. K. Jagannatham, "Cognitive decode-and-forward MIMO-RF/FSO cooperative relay networks," *IEEE Commun. Lett.*, vol. 21, no. 4, pp. 893–896, Apr. 2017.

[18] K. Kumar and D. K. Borah, "Quantize and encode relaying through FSO and hybrid FSO/RF links," *IEEE Trans. Veh. Technol.*, vol. 64, no. 6, pp. 2361–2374, Jun. 2015.

[19] I. Avram, N. Aerts, H. Bruneel, and M. Moeneclaey, "Quantize and forward cooperative communication: Channel parameter estimation," *IEEE Trans. Wireless Commun.*, vol. 11, no. 3, pp. 1167–1179, Mar. 2012.

[20] N. S. Ferdinand, N. Rajatheva, and M. Latva-Aho, "Effects of feedback delay in partial relay selection over Nakagami-m fading channels," *IEEE Trans. Veh. Technol.*, vol. 61, no. 4, pp. 1620–1634, May 2012.

[21] D. S. Michalopoulos, H. A. Suraweera, G. K. Karagiannidis, and R. Schober, "Amplify-and-forward relay selection with outdated channel estimates," *IEEE Trans. Commun.*, vol. 60, no. 5, pp. 1278–1290, May 2012.

[22] M. Soysa, H. A. Suraweera, C. Tellambura, and H. K. Garg, "Partial and opportunistic relay selection with outdated channel estimates," *IEEE Trans. Commun.*, vol. 60, no. 3, pp. 840–850, Mar. 2012.

[23] J. L. Vicario, A. Bel, J. Lopez-Salcedo, and G. Seco, "Opportunistic relay selection with outdated CSI: Outage probability and diversity analysis," *IEEE Trans. Wireless Commun.*, vol. 8, no. 6, pp. 2872–2876, Jun. 2009.

[24] G. K. Karagiannidis, "Performance bounds of multihop wireless communications with blind relays over generalized fading channels," *IEEE Trans. Wireless Commun.*, vol. 5, no. 3, pp. 498–503, Mar. 2006.

[25] D. B. Costa and S. Aissa, "Cooperative dual-hop relaying systems with beamforming over Nakagami-m fading channels," *IEEE Trans. Wireless Commun.*, vol. 8, no. 8, pp. 3950–3954, Aug. 2009.

[26] N. Yang, M. Elkashlan, J. Yuan, and T. Shen, "On the SER of fixed gain amplify-and-forward relaying with beamforming in Nakagami-m fading," *IEEE Commun. Lett.*, vol. 14, no. 10, pp. 942–944, Oct. 2010.

[27] S. Prakash and I. McLoughlin, "Performance of dual-hop multi-antenna systems with fixed gain amplify-and-forward relay selection," *IEEE Trans. Wireless Commun.*, vol. 10, no. 6, pp. 1709–1714, Jun. 2011.

[28] M. Lin, H. Wei, K. An, J. Ouyang, and Y. Huang, "Effect of imperfect channel state information and co-channel interferences on two-hop fixed gain amplify-and-forward relay networks with beamforming," *Int. J. Commun. Syst.*, vol. 28, no. 13, pp. 1921–1930, Sep. 2015.

[29] E. Costa and S. Pupolin, "M-QAM-OFDM system performance in the presence of a nonlinear amplifier and phase noise," *IEEE Trans. Commun.*, vol. 50, no. 3, pp. 462–472, Mar. 2002.

[30] T. Schenk, *RF Imperfections in High-rate Wireless Systems: Impact and Digital Compensation.* Dordrecht, Netherlands: Springer, 2008.

[31] C. Studer, M. Wenk, and A. Burg, "MIMO transmission with residual transmit-RF impairments," in *Proc. Int. ITG Workshop Smart Antennas (WSA)*, Feb. 2010, pp. 189–196.

[32] X. Zhang, M. Matthaiou, M. Coldrey, and E. Björnson, "Impact of residual transmit RF impairments on training-based MIMO systems," *IEEE Trans. Commun.*, vol. 63, no. 8, pp. 2899–2911, Aug. 2015.

[33] T. C. W. Schenk, E. R. Fledderus, and P. F. M. Smulders, "Performance analysis of zero-IF MIMO OFDM transceivers with IQ imbalance," *JCM*, vol. 2, no. 7, pp. 9–19, 2007.

[34] N. Maletic, M. Cabarkapa, and N. Neskovic, "Performance of fixed-gain amplify-and-forward nonlinear relaying with hardware impairments," *Int. J. Commun. Syst.*, vol. 30, no. 6, p. e3102, 2015.

[35] E. Bjornson, M. Matthaiou, and M. Debbah, "A new look at dual-hop relaying: Performance limits with hardware impairments," *IEEE Trans. Commun.*, vol. 61, no. 11, pp. 4512–4525, Nov. 2013.

[36] M. Matthaiou, A. Papadogiannis, E. Bjornson, and M. Debbah, "Two-way relaying under the presence of relay transceiver hardware impairments," *IEEE Commun. Lett.*, vol. 17, no. 6, pp. 1136–1139, Jun. 2013.

[37] J. Qi and S. Aïssa, "Analysis and compensation of power amplifier nonlinearity in MIMO transmit diversity systems," *IEEE Trans. Veh. Technol.*, vol. 59, no. 6, pp. 2921–2931, Jul. 2010.

[38] E. Balti, M. Guizani, B. Hamdaoui, and Y. Maalej, "Partial relay selection for hybrid FSO/RF systems with hardware impairments," in *Proc. IEEE Global Commun. Conf., Ad Hoc Sensor Netw. (Globecom AHSN)*, Washington, DC, USA, Dec. 2016, pp. 1–6.

[39] E. Balti, M. Guizani, and B. Hamdaoui, "Hybrid Rayleigh and double-Weibull over impaired RF/FSO system with outdated CSI," in *Proc. IEEE ICC Mobile Wireless Netw. (ICC MWN)*, Paris, France, May 2017, pp. 2525–2530.

[40] J. Li and J. Ilow, "Adaptive volterra predistorters for compensation of non-linear effects with memory in OFDM transmitters," in *Proc. 4th Annu. Commun. Netw. Services Res. Conf. (CNSR)*, May 2006, p. 4.

[41] F. H. Gregorio, "Analysis and compensation of nonlinear power amplifier effects in multi-antenna OFDM systems," Ph.D. dissertation, Dept. Elect, Commun., Eng., Helsinki Univ. Technol., Espoo, Finland, Nov. 2007.

[42] H. E. Rowe, "Memoryless nonlinearities with Gaussian inputs: Elementary results," *Bell Syst. Technol. J.*, vol. 61, no. 7, pp. 1519–1525, Sep. 1982.

[43] A. A. M. Saleh, "Frequency-independent and frequency-dependent non-linear models of TWT amplifiers," *IEEE Trans. Commun.*, vol. 29, no. 11, pp. 1715–1720, Nov. 1981.

[44] C. Rapp, "Effect of HPA-nonlinearity on 4-DPSK/OFDM-signal for a digital sound broadcasting system," in *Proc. 2nd Eur. Conf. Satellite Commun.*, vol. 2. Liege, Belgium, Oct. 1991, pp. 179–184.

[45] R. Zayani, R. Bouallegue, and D. Roviras, "Adaptive predistortions based on neural networks associated with Levenberg-Marquardt algorithm for satellite down links," *EURASIP J. Wireless Commun. Netw.*, vol. 2008, p. 2, Jan. 2008.

[46] J. P. Laico, H. L. McDowell, and C. R. Moster, "A medium power traveling-wave tube for 6,000-mc radio relay," *Bell Syst. Technol. J.*, vol. 35, no. 6, pp. 1285–1346, Nov. 1956.

[47] G. Santella and F. Mazzenga, "A hybrid analytical-simulation procedure for performance evaluation in M-QAM-OFDM schemes in presence of nonlinear distortions," *IEEE Trans. Veh. Technol.*, vol. 47, no. 1, pp. 142–151, Feb. 1998.

[48] N. Maletic, M. Cabarkapa, and N. Neskovic, "Performance of fixed-gain nonlinear relaying with hardware impairments," *Int. J. Commun. Syst.*, vol. 30, no. 6, p. e3102, 2015.

[49] H. Bouhadda, H. Shaiek, D. Roviras, R. Zayani, Y. Medjahdi, and R. Bouallegue, "Theoretical analysis of BER performance of nonlinearly amplified FBMC/OQAM and OFDM signals," *EURASIP J. Adv. Signal Process.*, vol. 2014, no. 1, p. 60, 2014.

[50] W. C. Jakes and D. C. Cox, Eds., *Microwave Mobile Communications*. Hoboken, NJ, USA: Wiley, 1994.

[51] I. S. Gradshteyn and I. M. Ryzhik, "Integral and series," *More Special Functions*, vol. 3, 7th ed. Orlando, FL, USA: Academic, 2007.

[52] C. Rapp, "Effects of HPA-nonlinearity on a 4-DPSK/OFDM-signal for a digital sound broadcasting system," in *Proc. Eur. Conf. Satellite Commun.*, Oct. 1991, pp. 179–184.

[53] G. Santella and F. Mazzenga, "A hybrid analytical-simulation procedure for performance evaluation in M-QAM-OFDM schemes in presence of nonlinear distortions," *IEEE Trans. Veh. Technol.*, vol. 47, no. 1, pp. 142–151, Feb. 1998.

[54] R. Zayani, R. Bouallegue, and D. Roviras, "Levenberg-marquardt learning neural network for adaptive predistortion for time-varying HPA with memory in OFDM systems," in *Proc. EUSIPCO*, 2008, pp. 1–5.

[55] *The Wolfram Functions Site*. [Online]. Available: http://functions.wolfram.com

[56] A. Erdelyi, W. Magnus, F. Oberhettinger, and F. G. Tricomi, *Tables of Integral Transforms* (Bateman Manuscript Project) New York, NY, USA: McGraw-Hill, 1954, vol. 1.

[57] A. Prudnikov and Y. A. Brychkov, "Integral and series," Dept. Math., Comput. Center USSR Acad. Sci., Moscow, Russia, Tech. Rep., 1990.

[58] F. Xu, F. C. M. Lau, and D. W. Yue, "Diversity order for dual-hop systems with fixed-gain relay under Nakagami fading channels," *IEEE Trans. Wireless Commun.*, vol. 9, no. 1, pp. 92–98, Jan. 2010.

[59] C. García-Corrales, F. J. Cañete, and J. F. Paris, "Capacity of $\kappa - \mu$ shadowed fading channels," *Int. J. Antennas Propag.*, vol. 2014, Jul. 2014, Art. no. 975109.

[60] H. Chergui, M. Benjillali, and S. Saoudi, "Performance analysis of project-and-forward relaying in mixed MIMO-pinhole and Rayleigh dual-hop channel," *IEEE Commun. Lett.*, vol. 20, no. 3, pp. 610–613, Mar. 2016.

[61] P. K. Mittal and K. C. Gupta, "An integral involving generalized function of two variables," *Proc. Indian Acad. Sci.-Section A*, vol. 75, no. 3, pp. 117–123, 1972.

[62] A. Soulimani, M. Benjillali, H. Chergui, and D. B. da Costa, "Performance analysis of M-QAM multihop relaying over mmWave Weibull fading channels," *CoRR*, Oct. 2016.

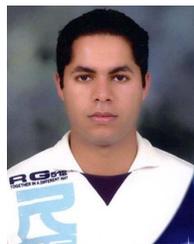

**Elyes Balti** (S'16) was born in Jendouba, Tunisia. He received the Diplôme d'Ingénieur degree in telecommunication engineering from the Ecole Supérieure des Communications de Tunis (Sup'-Com), Tunis, Tunisia, in 2013. He is currently pursuing the Ph.D. degree in electrical engineering with the University of Idaho, USA. His research interests include the modeling and performance analysis of optical wireless communication systems.

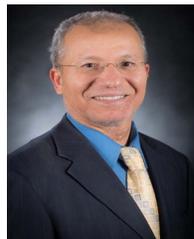

**Mohsen Guizani** (S'85–M'89–SM'99–F'09) received the B.S. (Hons.) and M.S. degrees in electrical engineering and the M.S. and Ph.D. degrees in computer engineering from Syracuse University, Syracuse, NY, USA, in 1984, 1986, 1987, and 1990, respectively. He served as the Associate Vice President of Graduate Studies, Qatar University, the Chair of the Computer Science Department, Western Michigan University, and the Chair of the Computer Science Department, University of West Florida. He also served in academic positions with the University of Missouri–Kansas City, the University of Colorado at Boulder, Syracuse University, and Kuwait University. He is currently a Professor and the ECE Department Chair with the University of Idaho, USA. He is the Author of nine books and over 450 publications in refereed journals and conferences. His research interests include wireless communications and mobile computing, computer networks, mobile cloud computing, security, and smart grid. He is a Senior Member of ACM. He also served as a member, the Chair, and the General Chair of a number of international conferences. He received the teaching award multiple times from different institutions and the best Research Award from three institutions. He was the Chair of the IEEE Communications Society Wireless Technical Committee and the Chair of the TAOS Technical Committee. He served as the IEEE Computer Society Distinguished Speaker from 2003 to 2005. He guest edited a number of special issues in IEEE journals and magazines. He currently serves on the editorial boards of several international technical journals and the Founder and the Editor-in-Chief of *Wireless Communications and Mobile Computing* journal (Wiley).